# Experiencing an elongated limb in virtual reality modifies the tactile distance perception of the corresponding real limb

François Le Jeune[1], Marco D'Alonzo[1], Valeria Piombino[1], Alessia Noccaro[1], Domenico Formica[1], Giovanni Di Pino[1]

[1] Research Unit of Neurophysiology and Neuroengineering of Human-Technology Interaction (NeXTlab), Università Campus Bio-Medico Di Roma, Via Álvaro Del Portillo 21, 00128 Rome, Italy

**Abstract**

In measurement, a reference frame is needed to compare the measured object to something already known. This raises the neuroscientific question of which reference frame is used by humans when exploring the environment. Previous studies suggested that, in touch, the body employed as measuring tool also serves as reference frame. Indeed, an artificial modification of the perceived dimensions of the body changes the tactile perception of external object dimensions. However, it is unknown if such a change in tactile perception would occur when the body schema is modified through the illusion of owning a limb altered in size. Therefore, employing a virtual hand illusion paradigm with an elongated forearm of different lengths, we systematically tested the subjective perception of distance between two points (tactile distance perception task, TDP task) on the corresponding real forearm following the illusion. Thus, TDP task is used as a proxy to gauge changes in the body schema. Embodiment of the virtual arm was found significantly greater after the synchronous visuo-tactile stimulation condition compared to the asynchronous one, and the forearm elongation significantly increased the TDP. However, we did not find any link between the visuo-tactile induced ownership over the elongated arm and TDP variation, suggesting that vision plays the main role in the modification of the body schema. Additionally, significant effect of elongation found on TDP but not on proprioception suggests that these are affected differently by body schema modifications. These findings confirm the body schema malleability and its role as reference frame in touch.

**Significance statement**

Evidence shows that humans use their body dimensions as reference frame to perceive the dimensions of the objects in the environment when using touch. We employed a modified version of the virtual hand illusion (VHI) to induce embodiment of fake elongated forearm. After the induction of the illusion, we tested if the perception of the distance between two points touching the real forearm changed. We show that the forearm elongation increases the perception of such distance. However, such elongation was not related to the embodiment level elicited by the VHI paradigm. These findings demonstrate the importance of the visual information in body schema modification, shedding new light on the relation between embodiment, body schema and world perception.

# 1    Introduction

Sensory feedback is exploited to experience the environment and guides physical interaction. In a nutshell, a parameter to be measured require unit of measure and a reference frame to which compare the measure. Especially, when the environment is experienced through our body (i.e., in somatosensation), it has been suggested that the reference frame we employ is the metric properties of our body (Proffitt and Linkenauger, 2013; Harris et al., 2015). This hypothesis is sound because it has been widely shown that the brain integrates a higher-order reconstruction of the body that is an implicit model of its metric properties (Longo and Haggard, 2010, 2012), namely the *body schema* (De Vignemont, 2010), and of the space where it interacts. This also means that change in body schema could affect the perception of the environment.

In touch, the perception of the distance between two stimulated points, i.e., tactile distance perception (TDP), depends on the part of the body which is stimulated. Such phenomenon, known as the Weber illusion (Weber, 1996), is certainly linked to the different tactile receptor density of different body parts, but there is more: for instance, a tactile receptor density variation of 340% results in a variation of only 30% in TDP between the palm of the hand and the forearm (Weinstein, 1968; Green, 1982). Hence, besides receptor density, a subsequent neural process must be involved in the Weber illusion. This process is a rescaling operation to compensate for the different receptor density, and it is likely to be based on a representation of the body parts in the body schema. Furthermore, since the body schema is continuously updated by multimodal sensory input (De Vignemont, 2010; Longo, 2010; Serino and Haggard, 2010; Romano et al., 2021), we hypothesize that a modification of the body schema achieved through multisensory feedback modulation would induce a consequent change of TDP. It has indeed been shown that by visually deforming the hand and forearm for about 1 hour, the TDP relative to the participant's forearm would change with respect to the not altered control part (Taylor-Clarke et al., 2004). In another experiment inspired by the Pinocchio illusion (Lackner, 1988), the proprioceptive illusion of an elongation of the index finger significantly increased the TDP on the same finger. The illusion of finger elongation was achieved by stimulating the spindles of the right arm biceps with a vibrator placed on the tendon (known as tendon-vibration illusion, Pinardi et al., 2020) while participants held their left index finger with their right arm (De Vignemont, 2005). Those studies confirmed that the perceived size of the body impacts on TDP, in line with the hypothesis that the body schema acts as reference frame for touch.

Another strategy to modify the perceived size of the body exploits the embodiment of fake hand through the rubber hand illusion (RHI) paradigm (Botvinick and Cohen, 1998) while the fake hand is placed in an artificial position, farther than the real hand, inducing the illusory feeling of having a longer arm (Armel and Ramachandran, 2003; Kilteni et al., 2012; Kalckert et al., 2019). Interestingly, previous works found correlations between the changes in perception of objects dimensions and the perceived embodiment, induced by a RHI paradigm, of a full-body or a rubber hand with rescaled dimensions, proving the existence of a link between a bodily illusion induced by the embodiment of external body part and the perception change (Van Der Hoort et al., 2011; Bruno and Bertamini, 2010).

However, to our knowledge, no previous study has investigated the possible change in *tactile distance perception* following a modification of the body schema through the embodiment of a fake limb. Thus, we used a RHI paradigm in virtual reality, a Virtual Hand Illusion (VHI), with an elongated virtual forearm to induce in participants the feeling of owning a distorted limb and investigated the evolution of the TDP on the corresponding real body part.

Following the findings of (Taylor-Clarke et al., 2004) and (De Vignemont et al, 2005) who, using *bodily illusion*, found that increased body size perception increases TDP, and in contrast with (Canzoneri et al, 2013) showing that *tool-use* reduces TDP, we hypothesized that the synchronous brush-stroking of a hand at the end of an elongated forearm would have increased the TDP on the corresponding real forearm and that the variation in TDP would have positively correlated with the forearm elongation and with the achieved level of embodiment of the hand. The asynchronous stimulation has been acquired as control condition.

## 2    Materials and Methods

### 2.1    Participants

Sixty-nine participants (33 females, 6 left-handed, aged 23.9±5.6) were enrolled in the study. For one of our analyses, participants were divided into three independent groups, depending on the first condition they would undergo (three conditions tested), thus each group containing twenty-three participants (see section 2.4 for detailed explanation). This number of enrolled participants per group has been based on the TDP task (TDPT) data distribution from the study of Taylor-Clarke and colleagues (Taylor-Clarke et al. 2004) to show a 7% mean shift in TDP between pre and post VHI, achieving an effect size of 0.32, a power superior to 0.8 and considering an independent t-test. All participants reported to have normal tactile sensation of the hand, forearm, and forehead, and normal or corrected-to-normal vision. All participants provided written informed consent before the experiment in accordance with the Declaration of Helsinki and following amendments. The experiment was conducted after approval of the Ethics Committee of [Author University].

### 2.2    Setup

Participant sat comfortably on a chair in front of a table placed within a 2.40m x 2.00m x 1.80m metallic structure. A large paper goniometer (58cm radius) was displayed upon the table. Participants visualized the immersive virtual environment through a virtual reality (VR) system (HTC Vive, HTC Corporation). They wore a VR headset (head mounted device – HMD) and the HMD movement was tracked by two infrared cameras (base-stations). To enhance the immersion of participants inside the virtual environment and their sense of agency over the virtual upper limb, the movements of their real left arm, forearm, hand, and fingers were tracked by motion capture systems. Arm and forearm movements were tracked with four infra-red cameras (Optitrack 13W, Natural Point, Inc) and reflective optical markers worn by the participant, whereas finger and hand movements were tracked by a dedicated infra-red motion tracking device attached on the HMD (Leap Motion, Ultraleap). We developed a VR environment (using the game engine Unity, version 2018.3.0, Unity Technologies) which replicates the lab room where the experiment was run, including the table and the chair. In the virtual environment, participants saw in a first-person perspective (1PP) their avatar's body (male or female) sitting in front of the virtual table. In the default position, both the participant and their avatar had the left forearm (palm down) on the table and the right arm alongside the body. The left forearm of the participant avatar could be elongated by different lengths (20cm or 40cm depending on the condition). The real experimenter, located in front of the participant, stimulated the participant's left hand index finger with a paintbrush. Both the experimenter and the paintbrush movement were replicated in the virtual environment (Fig. 1). The virtual avatars were created with the open-source software MakeHumanTM.

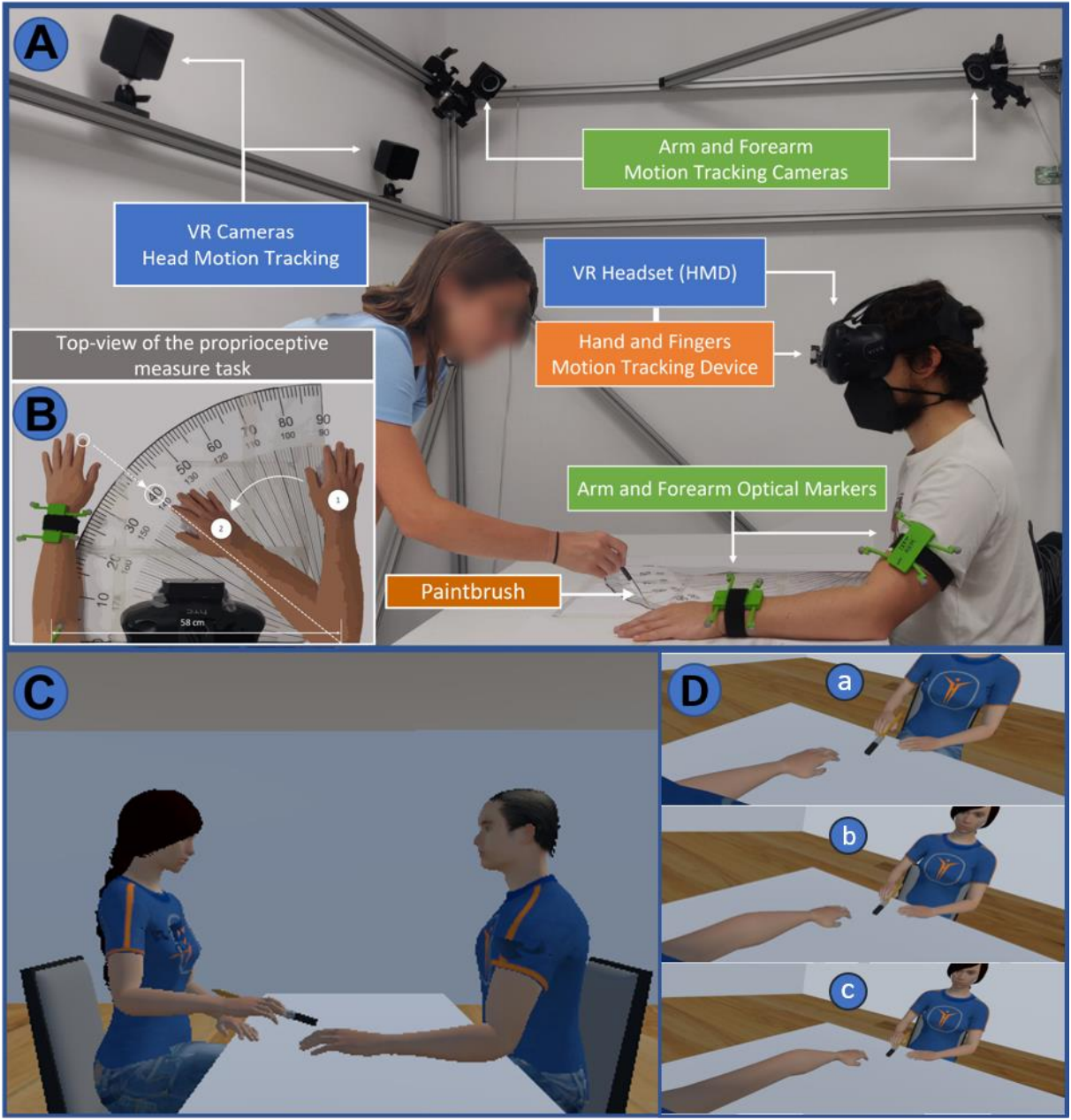

*Figure 1*: Real world and virtual experimental setup. **Panel A**: experimental setup with experimenter (left) and participant (right) in default position. **Panel B**: Proprioceptive drift measurement technique. **Panel C**: Virtual Environment – Sideview. Virtual experimenter (left) holding the paintbrush and participant avatar (right) in default position, without forearm elongation. **Panel D:** Virtual Environment – Participant's 1PP. Without forearm elongation (a), with 20cm elongation (b), with 40cm elongation (c).

## 2.3 Experiment

### 2.3.1 General procedure

We investigated the effects of a VHI with an elongated virtual forearm on TDP under three conditions. Synchronous VHI were performed with the virtual forearm elongated by 20cm (20S) or 40cm (40S), whereas an asynchronous VHI with a 20cm forearm elongation (20A) was used as control condition. We included only a single common control condition based on asynchronous tactile stimulation of the virtual hand. This experimental choice did not allow to apply a balanced factorial design to our study, but this choice has been made to reduce the duration of the experimental sessions for participant comfort and, consequently, reliability of the collected data and preventing the habituation to VHI (Bekrater-Bodmann et al., 2012; Convento et al., 2018).

Firstly, the virtual environment was turned pitch black, and participants underwent a preliminary task of TDP (pre-TDPT) to measure their baseline TDP (see section 2.3.2 for detailed information). Then, when VR was activated, participants were instructed to look at their surroundings (moving only their head) to familiarize with the virtual environment. To give participants agency over the avatar, they were asked to move their left arm for ninety seconds and then their hand and fingers for additional ninety seconds while looking at their virtual counterparts which moved accordingly.

Finally, participants were instructed to place back their left arm in the default position, and to keep it still until further notice. The VR environment was turned pitch black. Participants then performed a proprioceptive measure (pre-PM), in which they had to indicate the felt position of the tip of their left index finger (see section 2.3.2 for detailed information). The left virtual forearm was then elongated by 20cm or 40cm depending on the condition. The virtual environment was illuminated again, and participants were asked to look at and pay close attention to the (virtual) hand. To perform the VHI, the experimenter started the brush stroking synchronously or asynchronously (depending on the condition) with respect to the visual virtual brush stroking performed by the virtual experimenter on the left index finger of the participant avatar. During the asynchronous condition, the tactile stimulation on the real hand was delayed from the visual stimulation by 1s to ensure that both stimuli are not integrated (Maselli et al., 2016). Brush stroking lasted ninety seconds. Following the VHI, another proprioceptive measure (post-PM) was immediately performed, followed by a post-TDPT. Participants then were asked to answer a questionnaire evaluating the strength of the embodiment illusion elicited by the VHI. The latest steps, starting from the proprioceptive measure prior to the VHI, were repeated for every VHI condition, and conditions were tested in a pseudo-random order among participants. The whole procedure, including the preparation of the participant, lasted around one hour and half.

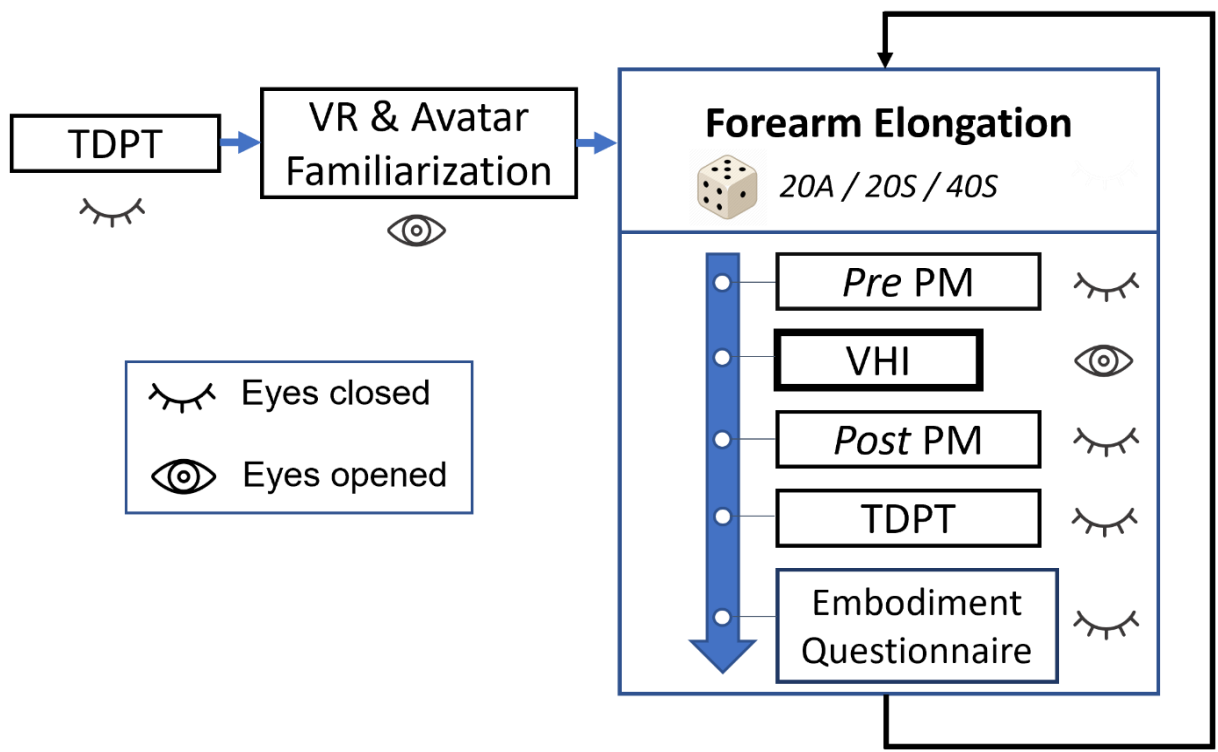

*Figure 2*: Experimental protocol of the experiment.

### 2.3.2 Measures

To measure the embodiment outcomes following the VHI, participants were asked to fill a self-evaluation questionnaire adapted from Botvinick and Cohen (1998) (D'Alonzo et al., 2019) to evaluate the strength of the ownership illusion over the virtual hand (Table 1). Three of the statements (Table 1: Q1, Q2 and Q3) were ownership-related and referred to the extent of sensory transfer into the virtual hand and its self-attribution during the VHI. The six other statements (Table 1: Q4, Q5, Q6, Q7, Q8 and Q9) served as control items to assess compliance, suggestibility, and placebo effect. For each statement, participants were asked to rate the extent to which these statements did or did not apply to their experience, by using a seven-point Likert scale. On this scale, -3 meant:" I am absolutely certain it did not apply", 0 meant:" uncertain whether it applied or not" and +3 meant:" I am absolutely certain it applied". The statements were presented to participants in a random order. The embodiment outcome of the VHI was taken from the questionnaire results and computed as the RHI Index. The RHI Index is defined as the difference between the mean score of the ownership statements and the mean score of the control statements (Abdulkarim and Ehrsson, 2016). The greater the RHI Index, the stronger the perceived embodiment.

| **VHI Embodiment Questionnaire** | | |
|---|---|---|
| **Embodiment statements** | Q1 | It seemed like I felt the touch of the paintbrush on the spot on which I was seeing the hand being touched. |
| | Q2 | It seemed like the touch I was feeling were due to the touch of the paintbrush on the hand I was seeing. |
| | Q3 | It felt like the hand I was seeing were my own hand. |
| **Control statements** | Q4 | It felt like my real hand were moving towards the hand I was seeing. |
| | Q5 | It felt like I had three arms or three hands. |
| | Q6 | It felt like the touch I was perceiving were coming from somewhere between my hand and the hand I was seeing. |
| | Q7 | It felt like my real hand were turning virtual. |
| | Q8 | It felt like the hand I was seeing was moving towards my real hand. |

| | Q9 | The hand I was seeing was starting to look like my real hand, in terms of shape, skin tone, freckles, or other characteristics. |

*Table 1*: Embodiment questionnaire with embodiment and control statements.

To measure the proprioceptive drift (PD) caused by the VHI (Botvinick and Cohen, 1998), participants were helped placing their right arm on the paper goniometer, parallelly to the left arm, in the starting position (90°). Participant forearms were always positioned on the extremities of the goniometer so that their elbows in-between distance would be of 580mm (see Fig. 1, Panel B). They were instructed to point with their right index finger towards the felt position of the tip of their left index finger by flexing the right forearm while keeping forearm, hand and finger along a straight line (Fig. 1, Panel B). We collected the corresponding angle and helped participants placing their right arm back alongside their body. Angular values were converted into the left elbow-index perceived distance expressed in millimeters, using a simple trigonometric function to obtain the elbow-index distance corresponding to the given angular value (θ) (eq. 1.1). The proprioceptive measure (PM) was performed right before (pre-PM) and after (post-PM) every VHI and the resulting PD was calculated as the difference between the post-PM and the pre-PM (eq. 1.2).

$$PM = 580 \times \tan(\theta) \tag{1.1}$$

$$PD = PM_{Post} - PM_{Pre} \tag{1.2}$$

To measure the TDP, participants underwent a TDPT. During the TDPT, blindfolded participants received fifty-six couples of tactile stimulation: one on the forearm (investigated TDP), and one on the forehead (used as reference TDP) in a random order. Tactile stimulations were performed using fork-like tools composed of two blunt tips with a specific distance between each other (Fig. 3). After each couple of stimulation, participants were asked to report in which of the two stimulations the distance between the tips of the forks was felt larger: they were instructed to answer "one" if the first stimulation was felt larger, "two" if it was the second. We recorded the corresponding body part on which the stimulation distance was felt larger. Distances between the tips of the forks were chosen based on a previous study using a similar task (Taylor-Clarke et al., 2004). In all couples of stimulations, we used a reference fork 45mm wide, and a fork with 30mm, 35mm, 40mm, 45mm, 50mm, 55mm or 60mm between-tips distance. The order, body part, and distance of the stimulations were randomized and balanced among the fifty-six couples of stimulations (twenty-eight unique couples of stimulations, each one performed twice, see Fig. 3).

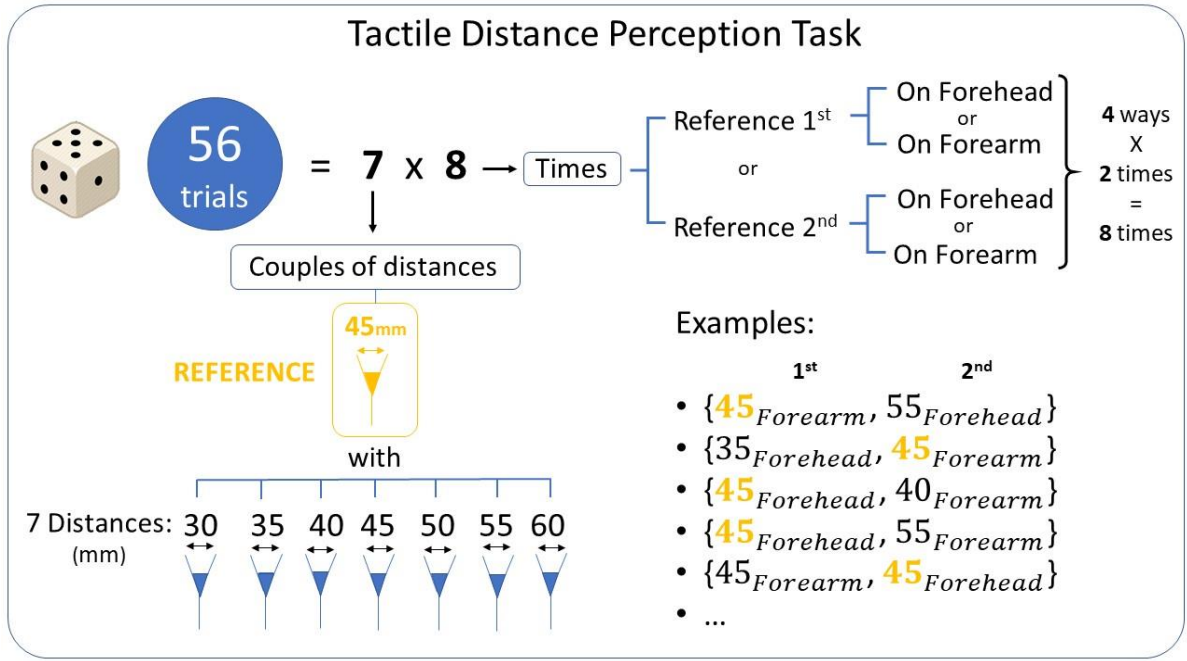

*Figure 3*: Tactile Distance Perception Task. 7 couples of distances tested in 4 different ways (28 different trials); each way repeated twice: 56 trials in total.

From the participants responses, we calculated the percentage of "forearm" answers (%ForeAns) for each difference (ΔL, in millimeters) between the length of the forearm stimulation (L$_{forearm}$) and the length of the forehead stimulation (L$_{forehead}$), positive values meaning bigger distance administered to the forearm (eq. 2).

$$\Delta L = L_{forearm} - L_{forehead} \tag{2}$$
$$\Delta L \in \{-15, -10, -5, 0, 5, 10, 15\}$$

The TDP was measured by the Point of Subjective Equality (PSE, in millimeters). As clearly defined in (Vidotto et al., 2019), "the point of subjective equality is any of the points along a stimulus dimension (here ΔL) at which a variable stimulus (here forearm stimulation distance) is judged by an observer to be equal to a standard stimulus (here forehead stimulation distance)". If the PSE is *positive*, it means that the stimulation distance on the forearm has to be *greater* (by "PSE" millimeters) than the stimulation distance on the forehead to be felt as equal to the stimulation distance on the forehead, and vice versa for negative PSE. Thus, a reduction of PSE means the actual forearm stimulation distance has to be "less larger" than the forehead stimulation than before to be felt equal to the forehead stimulation, hence meaning that is was *perceived* larger. To calculate the PSE, we proceeded as follows: for every participant and every TDPT (Pre, 20A, 20S, 40S) we plotted (Fig. 4) the %ForeAns (y-axis) in function of the ΔL (x-axis) as independent variable and fitted the data distribution with the following psychophysics sigmoid function (eq. 3.1):

$$P(\Delta L, PSE, EA) = \frac{100}{1 + \exp\left(-\frac{\Delta L - PSE}{0.5 \times EA}\right)} \tag{3.1}$$

adapted from (Di Pino et al., 2020) where P is the probability (expressed as a percentage) of feeling the larger stimulation of the forearm. From the plotted curve (Fig. 4), the PSE can be defined as:

$$PSE = \Delta L|_{P_{50\%}} \tag{3.2}$$

It represents the ΔL value of the point in which the curve corresponds to the P = 50% value, i.e., the ΔL value for which the participant has the same probability to perceive the larger stimulation on the forearm as on the forehead (perceived equality of distances). EA is the esteem accuracy and represents the ΔL value of the point in which the line tangent to the curve at the point of coordinates (PSE, 50%) reaches the value P = 100%, subtracted by the PSE value. It is the inverse of twice the slope of the curve at (PSE, 50%). Considering that we expected a change in TDP and not in tactile accuracy, the EA was not further considered in the study.

PSE values were obtained from the fittings results. The goodness-of-fit (R-squared) of all analyzed participants was above 0.6 (moderate effect size, Moore et al., 2013)

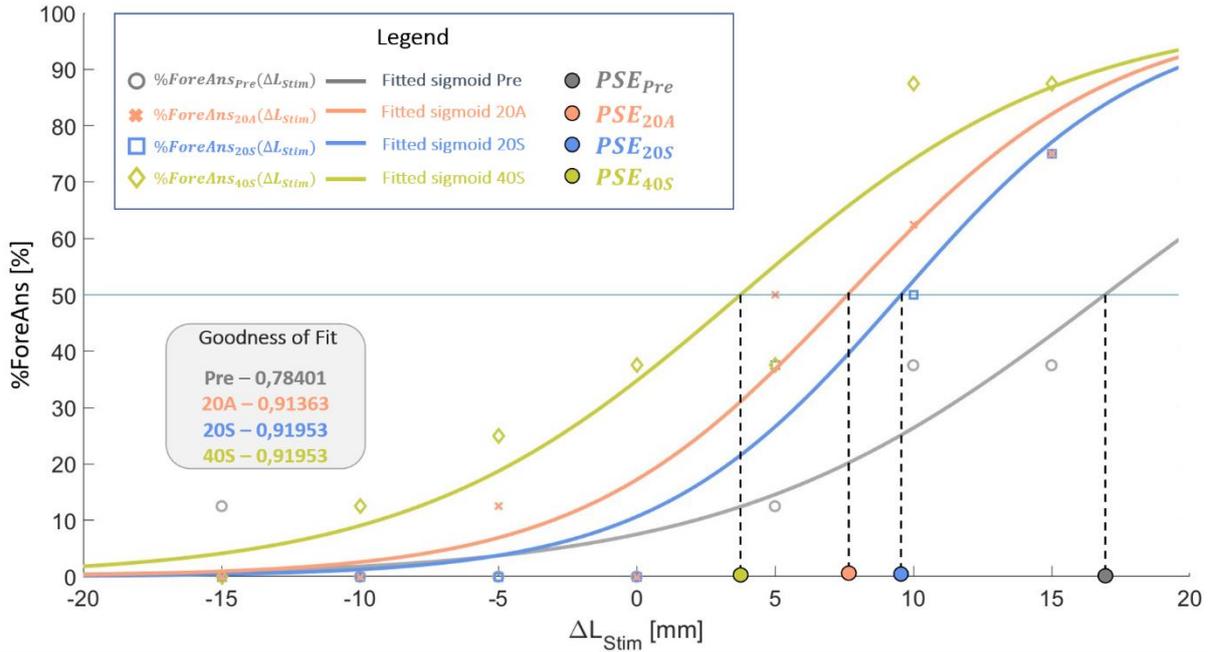

*Figure 4*: Example of a typical TDPT results plot for one subject. Raw results are the percentages of "forearm" answers in function of stimulation distance difference between the stimulation on the forearm and the stimulation on the forehead (ΔL). They are identified with grey empty circles (*Pre*), red crosses (*20A*), blue squares (*20S*), and green diamonds (*40S*). Continuous curves are their color-corresponding fitted sigmoid curves with their goodness-of-fit values specified in the top-left-hand corner. Filled circles on the abscise indicate the graphical definition of the PSE.

Then, we computed for each condition the variation of PSE, i.e., ΔPSE, the difference between the post-PSE (after the VHI) and the pre-PSE (baseline PSE for the first tested condition and PSE after the previous condition for following conditions).

## 2.4  Data analysis

We firstly planned a general within-subjects analysis (i.e., repeated measures). We considered the possibility that by undergoing different conditions successively without a pause between them, one condition could have an influence on the successive one and that by repeating the VHI, its intensity might decrease. Therefore, to investigate the *order* effect, i.e. the order in which the conditions were presented to participants on our outcomes and to counter-balance the lack of a 2-by-2 factorial design (no 40A condition), we ran a linear mixed model (LMM) considering the effects of *synchronicity*, *elongation*, and *order*. We expected an influence of the *order* effect, i.e. a predominant effect of the first condition presented to participants, and to discard its aforementioned effects from the analysis, a between-subject analysis was planned regarding the data relative to the first condition presented to participants only. Thus, three independent groups (of 23 subjects each) were formed depending on the first condition tested. In both analyses, the significance threshold was set to p-values lower than 0.05 for all statistical tests. Correlation analyses were performed between all three outcomes and for each outcome we pooled together data from all conditions. All statistical tests were conducted with JASP (JASP Team, 2023).

### 2.4.1 General analysis

For each condition, all data (RHI Index, PD and ΔPSE) were distributed normally (Shapiro-Wilk test, p>0.05). We used a LMM on each of our outcomes. For each LMM, we introduced the outcome as the dependent variable, included fixed effects of *synchronicity*, *elongation* and *order*, and included subjects as random effect. Post-hoc tests were performed through the study of the contrasts (Bonferroni corrected) between the conditions of interest. We tested the significance of the variations of PD and ΔPSE with respect to 0 using one-sample Student t-tests. To investigate any difference of distribution in *pre* measures between conditions (20S vs 20A and 40S vs 20S) for the PD and ΔPSE i.e., pre-PM and pre-PSE, we used Student's paired t-tests on pre-PM data (normally distributed among conditions, Bonferroni corrected for two comparisons) and Wilcoxon's signed-ranked t-tests on pre-PSE data (non-normally distributed among conditions, Bonferroni corrected for two comparisons). For correlation analyses, the data were pooled across conditions. Spearman's ρ was calculated between RHI Index and PD, and between RHI Index and ΔPSE (RHI Index data non-normally distributed) and Pearson's *r* was calculated between PD and ΔPSE.

### 2.4.2 First condition analysis

Among each group, all data (RHI Index, PD, ΔPSE) were normally distributed (Shapiro-Wilk test, p>0.05). We used pre-planned independent Student's t-tests to evaluate the effects of *synchronicity* (20S vs 20A) and *elongation* (40S vs 20S) on RHI Index, PD and ΔPSE (Bonferroni corrected for two comparisons). We tested the significance of the variations of PD and ΔPSE with respect to 0 using one-sample Student t-tests. Independent Student's t-tests were used to investigate any difference of distribution in pre-PM and pre-PSE data between conditions (normally distributed among conditions, Bonferroni corrected for two comparisons). For correlation analyses, Pearson's *r* was calculated between all outcomes.

**Statistical Table 1**

| Outcome | Effect | df | F | p-values |
|---|---|---|---|---|
| RHI Index | Synchronicity | 1, 134.00 | 93.928 | < 0.001*** |
| RHI Index | Elongation | 1, 134.00 | 0.004 | 0.947 |
| RHI Index | Order | 2, 134.00 | 7.247 | 0.001** |
| PD | Synchronicity | 1, 160 | 3.847 | 0.052 |
| PD | Elongation | 1, 160 | 0.837 | 0.362 |
| PD | Order | 2, 160 | 3.944 | < 0.021* |
| ΔPSE | Synchronicity | 1, 190 | 1.517 | 0.220 |
| ΔPSE | Elongation | 1, 190 | 5.223 | 0.023* |
| ΔPSE | Order | 2, 190 | 12.015 | < 0.001*** |

*Table 2*: Statistical table of the **Linear Mixed Model** used for **the general analysis** with corresponding degrees of freedom (df), F-statistic values (F), and p-values (asterisks meanings: $p<0.05$: *; $p<0.01$: **; $p<0.001$: ***).

**Statistical Table 2**

| Outcome | Comparison | Estimate | SE | 95% CI (Lower) | 95% CI (Upper) | z | p-value |
|---|---|---|---|---|---|---|---|
| RHI Index | 20S vs 20A | 5.237 | 0.540 | 4.178 | 6.296 | 9.692 | <0.001*** |
| RHI Index | 40S vs 20A | 5.201 | 0.541 | 4.141 | 6.261 | 9.619 | <0.001*** |
| RHI Index | 1st vs 2nd | 2.439 | 0.720 | 1.027 | 3.851 | 3.386 | 0.004** |
| RHI Index | 1st vs 3rd | 2.307 | 0.721 | 0.894 | 3.720 | 3.200 | 0.008** |
| RHI Index | 2nd vs 3rd | -0.133 | 0.721 | -1.546 | 1.280 | -0.184 | 1.000 |
| PD | 1st vs 2nd | 207.965 | 75.085 | 60.800 | 355.130 | 2.770 | 0.017* |
| PD | 1st vs 3rd | 134.197 | 75.085 | -12.968 | 281.361 | 1.787 | 0.222 |
| PD | 2nd vs 3rd | -73.768 | 75.085 | -220.933 | 73.396 | -0.982 | 0.978 |
| ΔPSE | 40S vs 20A | -2.186 | 2.069 | -6.242 | 1.870 | -1.056 | 1.000 |
| ΔPSE | 40S vs 20S | -4.729 | 2.069 | -8.785 | -0.673 | -2.285 | 0.111 |
| ΔPSE | 1st vs 2nd | -11.207 | 2.753 | -16.603 | -5.810 | -4.070 | <0.001*** |
| ΔPSE | 1st vs 3rd | -12.136 | 2.759 | -17.544 | -6.728 | -4.398 | <0.001*** |
| ΔPSE | 2nd vs 3rd | -0.929 | 2.759 | -6.337 | 4.478 | -0.337 | 1.000 |

*Table 3*: Statistical table of the contrasts analyses from the LMM (Table 2) with corresponding estimates, standard errors (SE), confidence intervals (95% CI lower and upper), z-statistics values (z) and p-values (asterisks meanings: p<0.05: *; p<0.01: **; p<0.001: ***)

**Statistical Table 3**

| Outcome | Condition | df | Cohen's d | SE Cohen's d | 95% CI (Lower) | 95% CI (Upper) | t | p-value |
|---|---|---|---|---|---|---|---|---|
| PD | 20A | 54 | 0.004 | 0.135 | -0.260 | 0.268 | 0.030 | 0.976 |
| PD | 20S | 54 | 0.420 | 0.141 | 0.142 | 0.694 | 3.112 | 0.003** |
| PD | 40S | 54 | 0.456 | 0.142 | 0.176 | 0.732 | 3.382 | 0.001** |
| ΔPSE | 20A | 64 | -0.213 | 0.125 | -0.458 | 0.034 | -1.717 | 0.091 |
| ΔPSE | 20S | 64 | 0.039 | 0.124 | -0.204 | 0.282 | 0.316 | 0.753 |
| ΔPSE | 40S | 64 | -0.285 | 0.127 | -0.532 | -0.036 | -2.295 | 0.025* |

*Table 4*: Statistical table of the one-sample Student t-tests analyses from the general analysis with corresponding degrees of freedom (df), Cohen'd (as effect size), SE (standard error) of Cohen's d, confidence intervals of Cohen's d (95% CI lower and upper), t-statistics (t) and p-values (asterisks meanings: $p<0.05$: *; $p<0.01$: **; $p<0.001$: ***).

**Statistical Table 4**

| Outcome | Effect (Comparison) | df | Cohen's d | SE Cohen's d | 95% CI (Lower) | 95% CI (Upper) | t | p-value |
|---|---|---|---|---|---|---|---|---|
| RHI Index | Synchronicity (20S vs 20A) | 44 | 1.024 | 0.331 | 0.403 | 1.635 | 3.473 | 0.002** |
| RHI Index | Elongation (40S vs 20S) | 44 | 0.179 | 0.296 | -0.757 | 0.402 | 0.606 | 1.000 |
| PD | Synchronicity (20S vs 20A) | 42 | 0.232 | 0.304 | -0.362 | 0.824 | 0.770 | 0.892 |
| PD | Elongation (40S vs 20S) | 43 | 0.298 | 0.302 | -0.884 | 0.292 | -0.999 | 0.646 |
| ΔPSE | Synchronicity (20S vs 20A) | 43 | 0.029 | 0.298 | -0.556 | 0.613 | 0.096 | 1.000 |
| ΔPSE | Elongation (40S vs 20S) | 41 | -0.778 | 0.327 | -1.395 | -0.153 | -2.551 | 0.030* |

*Table 5*: Statistical table of the **independent Student's t-tests of the first condition analysis** with corresponding degrees of freedom (df), Cohen'd (as effect size), SE (standard error) of Cohen's d, confidence intervals of Cohen's d (95% CI lower and upper), t-statistics (t) and p-values (asterisks meanings: $p<0.05$: *; $p<0.01$: **; $p<0.001$: ***).

**Statistical Table 5**

| Outcome | Condition | df | Cohen's d | SE Cohen's d | 95% CI (Lower) | 95% CI (Upper) | t | p-value |
|---|---|---|---|---|---|---|---|---|
| PD | 20A | 22 | 0.571 | 0.225 | 0.123 | 1.007 | 2.737 | 0.012* |
| PD | 20S | 22 | 0.740 | 0.235 | 0.270 | 1.196 | 3.547 | 0.002** |
| PD | 40S | 22 | 0.650 | 0.229 | 0.193 | 1.095 | 3.116 | 0.005** |
| ΔPSE | 20A | 22 | -0.544 | 0.223 | -0.977 | -0.100 | -2.608 | 0.016* |
| ΔPSE | 20S | 21 | -0.488 | 0.226 | -0.926 | -0.040 | -2.288 | 0.033* |
| ΔPSE | 40S | 20 | -1.360 | 0.303 | -1.949 | -0.753 | -6.232 | <0.001*** |

*Table 6*: Statistical table of the **one-sample Student t-tests of the first condition analysis** with corresponding degrees of freedom (df), Cohen'd (as effect size), SE (standard error) of Cohen's d, confidence intervals of Cohen's d (95% CI lower and upper), t-statistics (t) and p-values (asterisks meanings: $p<0.05$: *; $p<0.01$: **; $p<0.001$: ***).

**Statistical Table 6**

| Analysis | Outcomes | Data distrib. | Coef. | r or ρ | Fisher's z | 95% CI (Lower) | 95% CI (Upper) | p-value |
|---|---|---|---|---|---|---|---|---|
| General | RHI Index - PD | Non-normal (RHI Index) | ρ | 0.187 | 0.189 | 0.035 | 0.331 | 0.016* |
| General | RHI Index - ΔPSE | Non-normal (RHI Index) | ρ | -0.125 | -0.126 | -0.267 | 0.021 | 0.094 |
| General | PD - ΔPSE | Normal | r | -0.056 | -0.056 | -0.213 | 0.103 | 0.488 |
| 1st condition | RHI Index – PD | Normal | r | 0.054 | 0.054 | -0.189 | 0.290 | 0.666 |
| 1st condition | RHI Index - ΔPSE | Normal | r | -0.225 | -0.229 | -0.443 | 0.018 | 0.069 |
| 1st condition | PD - ΔPSE | Normal | r | 0.029 | 0.029 | -0.219 | 0.273 | 0.821 |

*Table 7*: Statistical table of the **correlation analyses** of both general and first-condition analyses with corresponding Pearson's r values or Spearman's ρ values, Fisher's z (as effect size), confidences intervals (95% CI lower and upper) and p-values (asterisks meanings: $p<0.05$: *; $p<0.01$: **; $p<0.001$: ***).

# 3 Results

A summary of the results can be found in Table 2 to Table 7 with corresponding statistical values.

## 3.1 General analysis

Considering the RHI Index, there was a significant effect of *synchronicity* (Fig.5, top-left graph; df = 1, 134.00; F = 93.928; p<0.001). Analysis of contrasts (Bonferroni corrected) showed that the score of the 20cm synchronous VHI (20S) resulted significantly greater than the asynchronous control condition (20A) (z = 9.692, p<0.001), and that so did the 40cm synchronous VHI (40S) (z = 9.619, p<0.001). The *elongation* effect was not found significant (Fig.5, top-right graph; df = 1, 134.00; F = 0.004; p = 0.947). However, the *order* effect was found significant (Fig. 6, top graph; df = 2, 134.00; F = 7.247; p = 0.001). The analysis of contrasts showed that the VHI score of the first condition presented to participant was significantly greater than both the second one (z = 3.386, p = 0.004) and the third one (z = 3.200, p = 0.008). No significant difference was found in VHI score between the second and third condition (z = -0.184; p = 1.000).

The *synchronicity* effect on PD was found very close to being significant (Fig.5, middle-left graph; df = 1, 160; F = 3.847; p = 0.052). The *elongation* effect was not found significant on PD (Fig.5, middle-right graph; df = 1, 160; F = 0.837; p = 0.365). Nevertheless, PD resulted significantly greater than 0 following the synchronous conditions (20S: df = 54, t = 3.035, p = 0.004; 40S: df = 54, t=3.499, p<0.001) but not following the asynchronous one (20A: df = 54, t = 0.118, p = 0.906) and no significant difference of PD pre-measures (pre-PM) was found neither between 20S and 20A (df = 54, t = -0.940, p = 0.704) nor between 40S and 20S (df = 54, t = 0.583, p = 1.000). A significant effect of *order* on PD (Fig. 6, middle graph; df = 2, 160; F = 3.944; p = 0.021) was highlighted. Contrasts revealed that the PD elicited by the first condition presented to participants was significantly greater than the PD elicited by the second condition (z = 2.770; p = 0.017) but not significantly greater than the PD elicited by the third condition (z = 1.787; p = 0.222). No significant difference was found in PD between the second and third condition (z = -0.982; p = 0.978).

No significant effect of *synchronicity* was found on ΔPSE (Fig.5, bottom-left graph; df = 1, 190; F = 1.517; p = 0.220). There was a significant effect of *elongation* (Fig.5, bottom-right graph; df = 1, 190; F = 5.223; p = 0.023). The contrasts however showed no significant difference in ΔPSE neither between 40S and 20A (z = -1.056, p = 1.000), nor between 40S and 20S (z = -2.285, p = 0.111). ΔPSE was found significantly lower than 0 following the 40S condition but not for other conditions (20A: df = 64, t = -1.717, p = 0.091; 20S: df = 64, t = 0.316, p = 0.753; 40S: df = 64, t = -2.295, p = 0.025) and no significant difference of PSE pre-measures (pre-PSE) was found neither between 20S and 20A (z = -1.101, p = 0.544) nor between 40S and 20S (z = 0.186, p = 1.000). Nevertheless, there was a significant effect of *order* on ΔPSE (Fig. 6, bottom graph; df = 2, 192; F = 12.015; p < 0.001). The contrasts showed that the ΔPSE was significantly lower after the first condition than both the second condition (z = -4.070; p < 0.001) and the third condition (z = -4.398; p < 0.001). No significant difference was found in ΔPSE between the second and third condition (z = -0.337; p = 1.000). A significant correlation was found between RHI Index and PD (Fig. 8.A, top graph; ρ = 0.187, p = 0.016), but neither between RHI and ΔPSE (Fig. 8.A, bottom-left graph; ρ = -0.125, p = 0.094) nor between PD and ΔPSE (Fig. 8.A, bottom-right graph; r = -0.056, p = 0.488).

### 3.2   First condition analysis

The *synchronicity* effect was found significant on the RHI Index (Fig. 7, top-left graph). Indeed, the score of the synchronous VHI resulted significantly greater than the asynchronous control condition (20S vs 20A: t = 3.473, p = 0.002) and no significant effect of *elongation* was found on the RHI Index (Fig. 7, top-right graph; 20S vs 40S: t = 0.606, p = 1.000).

No significant difference of PD pre-measures was found between conditions (20S vs 20A: t = -0.004, p = 1.000 and 40S vs 20S: t = 1.063, p = 0.588) and all conditions were found to elicit a PD significantly higher than 0 (Fig. 7, middle line graphs; 20A: t = 2.737, p = 0.012; 20S: t = 3.547, p = 0.002; 40S: t = 3.116, p = 0.005). However, no significant effect of *synchronicity* nor *elongation* was found on PD (20S vs 20A: t = 0.770, p = 0.892; 40S vs 20S: t = -0.999, p = 0.646).

No significant difference of PSE pre-measures (pre-PSE) was found between conditions 20S and 20A (t=1.463, p=0.302) nor between 40S and 20S (t=1.029, p=0.620). All conditions were found to elicit a ΔPSE significantly lower than 0 (Fig. 7, bottom line graphs; 20A: t=-2.608, p=0.016; 20S: t=-2.288, p=0.033; 40S: t=-6.232, p<0.001). No significant effect on *synchronicity* was found on ΔPSE (20S vs 20A: t=0.096, p=1.000), but the *elongation* effect was significantly present (40S vs 20S: t=-2.551, p=0.030). The correlation between RHI Index and ΔPSE was found close to significance (Fig. 8.B , bottom-left graph; r=-0.225, p=0.069) and no significant correlation was found neither between RHI Index and PD (Fig. 8.B, top graph; r=0.054, p=0.666) nor between PD and ΔPSE (Fig. 8.B, bottom-right graph; r=0.029, p=0.821).

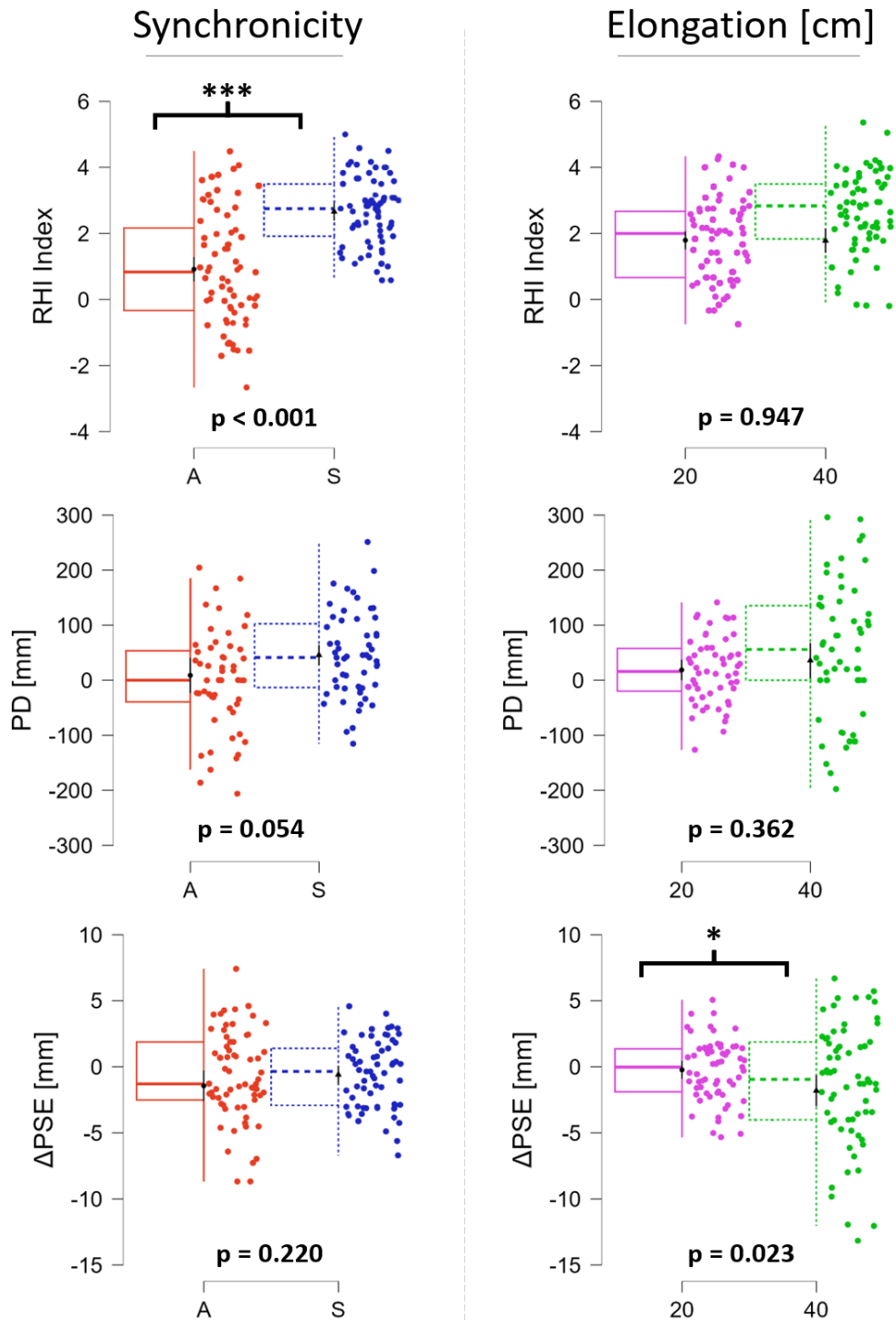

*Figure 5*: Box-jitter plots showing results of the **general analysis** (LMM) on the *synchronicity* effect (left) and *elongation* effect (right) on the RHI Index (top), PD (middle) and ΔPSE (bottom). P-values refer to the specific effect on the specific outcome. Asterisks meanings: p<0.05: *; p<0.01: **; p<0.001: ***.

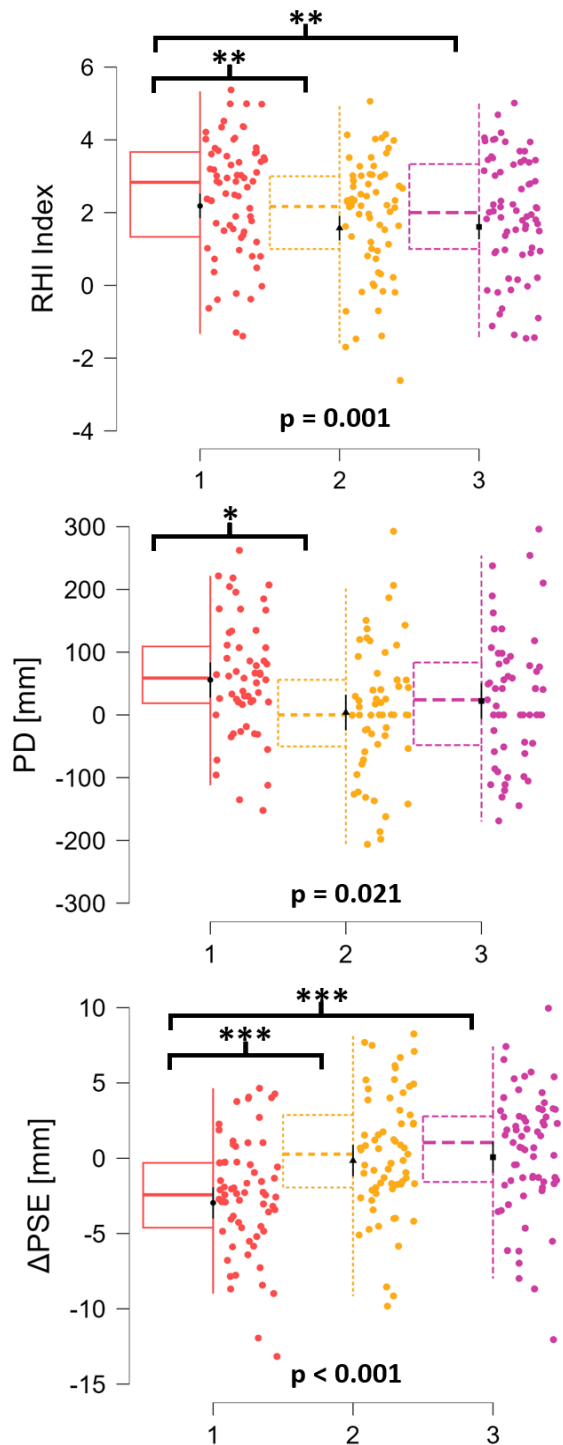

*Figure 6*: Box-jitter plots showing results of the **general analysis** (LMM) on the *order* of the condition presented to participants on the RHI Index (top), PD (middle) and ΔPSE (bottom). P-values refer to the *order* effect on the specific outcome. Asterisks meanings: p<0.05: *; p<0.01: **; p<0.001: ***.

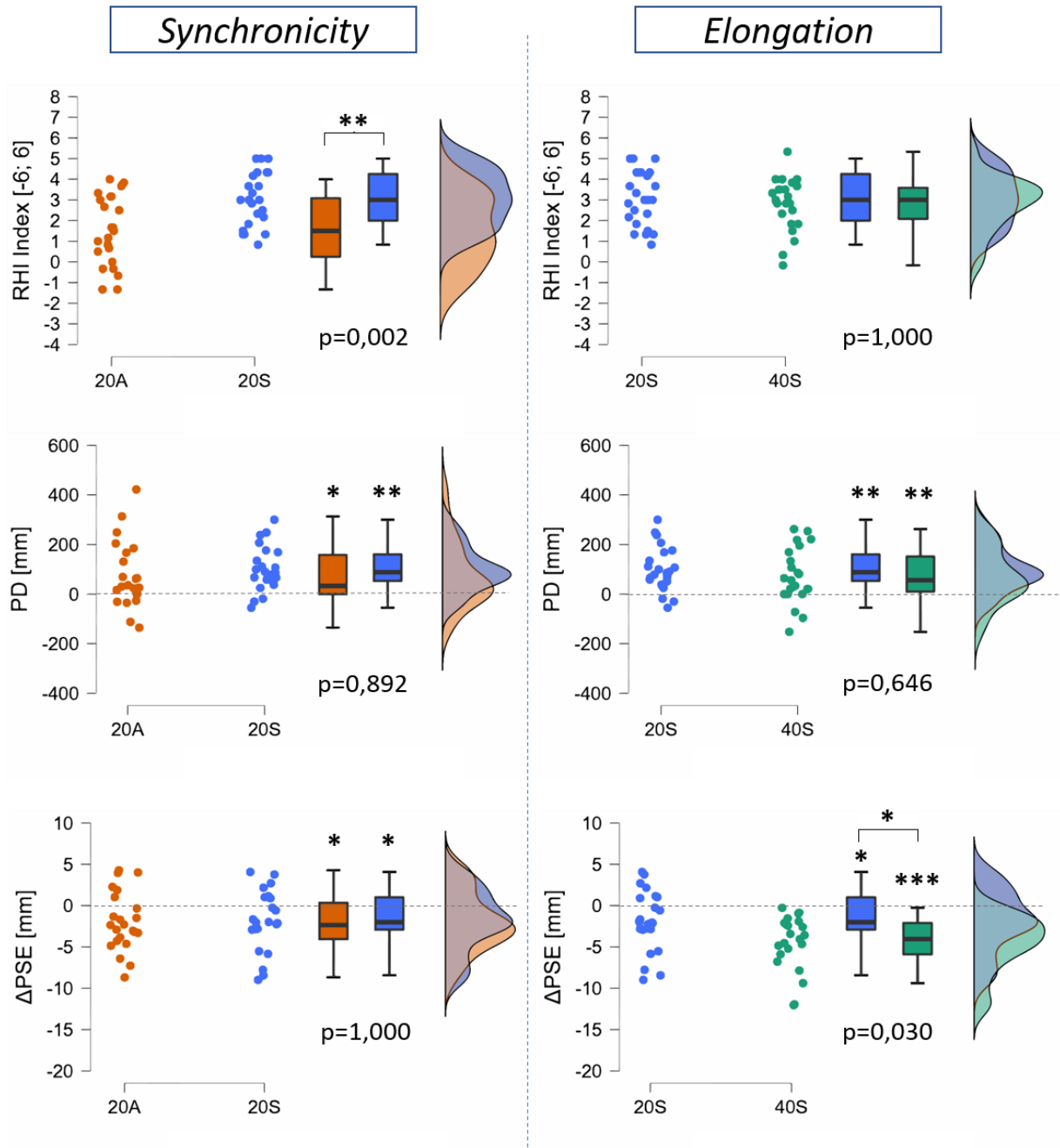

*Figure 7*: Raincloud plots showing results of the **first condition analysis** on the *synchronicity* effect (left) and *elongation* effect (right) on the RHI Index (top), PD (middle) and ΔPSE (bottom). Asterisks on top of a singular box plot (without brackets) indicate the significance level of the difference from 0. Asterisks on top of brackets indicate the significance level of the difference between the indicated conditions and the corresponding p-value is displayed below the box plots. Asterisks meanings: $p<0.05$: *; $p<0.01$: **; $p<0.001$: ***.

## General analysis

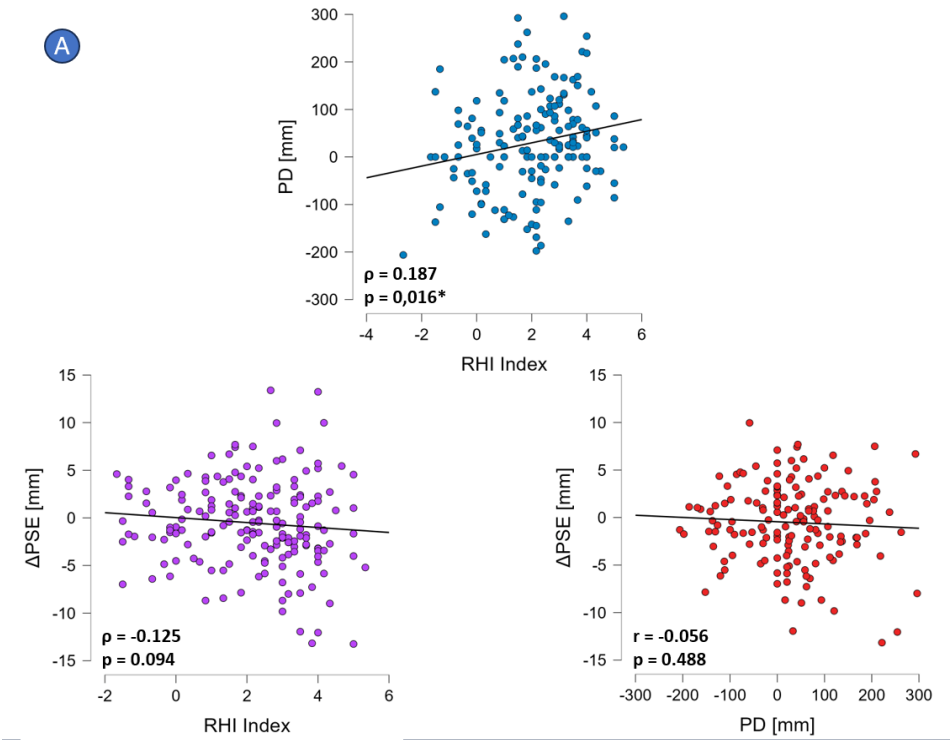

## 1st condition analysis

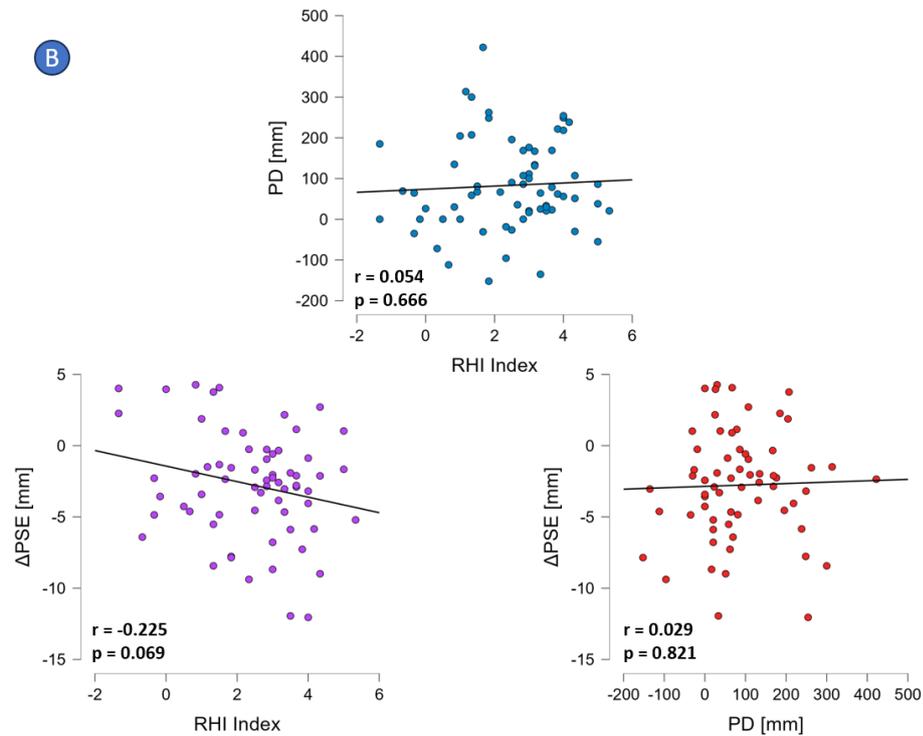

*Figure 8*: Correlations plots for the general analysis (A) and for the analysis on the first condition (B): RHI Index and PD (top), RHI Index and ΔPSE (bottom-left), PD and ΔPSE (bottom-right). ρ and *r* represent Spearman's ρ and Pearson's *r*, respectively. Asterisk meanings: p<0.05: *.

# 4 Discussion

The aim of the study was to understand whether the change in TDP following a bodily illusion demonstrated by previous studies (Taylor-Clarke et al. 2004, De Vignemont et al. 2005) could take place through *embodiment* over a fake limb. We used a VHI paradigm with an elongated forearm in a 1PP virtual environment to induce the bodily illusion of owning an elongated forearm and assessed the resulting change in subjective perception of distance between two simultaneously touched points using a TDPT. Through a linear mixed model, we investigated on each of our outcomes (RHI Index, PD, ΔPSE) the effects of the *synchronicity* of the VHI, of the forearm *elongation*, and of the *order* in which the conditions (20A, 20S, and 40S) were presented to participants. We also investigated the correlation between different aspects of the elongated forearm embodiment and the changes in TDP. The embodiment illusion proved effective, as the effect of *synchronicity* was found significant with participants perceiving a significantly higher level of ownership (from the RHI Index) after the synchronous VHI (20S and 40S) compared to the asynchronous condition (20A). Our experiment confirms previous findings on the effectiveness of synchronous visuo-tactile stimulation in eliciting ownership over a virtual limb (Slater et al. 2008, Pyasik et al. 2020). Besides, no effect of the forearm elongation on the perceived ownership was highlighted as the *elongation* effect did not result significant, meaning that both elongations resulted in a similar embodiment level. This comes in line with previous results showing that it is possible to embody a fake hand placed farther than the real one along the distal plane with a RHI or VHI paradigm (Armel and Ramachandran, 2003; Kilteni et al., 2012; Kalckert et al., 2019). More specifically, previous study (Kilteni et al., 2012) showed a steady ownership level with a virtual forearm of one, two and three times the normal forearm length. Considering that the elongation magnitudes (20 and 40 cm elongations) of our study are comprised in these lengths (forearm length of participants: 25,77 ± 2.22 cm, thus a times-three forearm length would be equivalent to a 50 cm elongation), our results confirm the finding of that study.

From the general analysis, synchronous conditions (20S and 40S) caused a significant drift of the perceived position of the hand from the one perceived prior to the VHI towards the position of the virtual hand visualized during the VHI, whereas the asynchronous condition (20A) did not. This result would hint towards an effect of the synchronicity of the VHI on the PD with a higher PD due to the synchronous VHI with respect to the asynchronous VHI. Nevertheless, although very close to significance, the VHI *synchronicity* was not found to influence the PD, neither was the forearm *elongation* magnitude. Mixed results were found considering the change in TDP. On the one hand, as no significant effect of synchronicity was found, the VHI synchronicity did not prove to influence the change in TDP. This comes in opposition with our first hypothesis on the causal link between the synchronous visuo-tactile stimulation and the increase in TDP (decrease of the PSE). On the other hand, as shown by the linear mixed model analysis, the *elongation* effect proved significant on ΔTDP (significant main effect of the elongation). Furthermore, the greater elongation condition (40S) significantly increased the TDP (significantly decreased the PSE) with respect to the pre-VHI perception, whereas smaller elongation conditions (20A and 20S) did not. This suggests the presence of an effect of the *elongation* magnitude of the virtual forearm, with greater increases in TDP for greater elongations.

Different factors could have affected our results. Testing multiple repetitions of the RHI paradigm successively could have decreased participant sensitivity to the illusion. Additionally, participants underwent all conditions successively without removing the headset nor moving their left arm. Therefore, no proper "reset" of the experimental conditions took place between the different conditions, and thus the effects of one condition might have been transferred to the successive condition. The analysis of the effect of the *order* in which the conditions were presented to the participants proves that those hypotheses are sound. Indeed, for all three outcomes, the *order* effect was found significant, and more specifically, the contrasts showed that all outcomes where significantly greater after the first condition than after the second and third condition (except for the third condition of the PD). Considering the fact that the aforementioned ~~These~~ issues could have affected the second and third conditions undergone by participants but not of the first one, and taking into account the crucial effect of the *order* in which the conditions were presented to the participants, the between-subjects analysis on the first condition tested appears greatly relevant.

Regarding the perceived embodiment of the virtual hand through the VHI, the analysis on the first condition gave identical results as the previous analysis. However, the analysis on the first condition revealed that, after all VHI conditions, participants perceived the position of their real hand significantly drifted towards the position of the virtual hand (PD), independently from the VHI *synchronicity* or the magnitude of *elongation*. Similarly to PD findings, ΔPSE analysis showed that participants perceived an increase in TDP on the real forearm after all elongated VHI conditions, and that, as previously suggested, a greater virtual forearm elongation results in a greater augmentation in TDP (ΔPSE due to 40S significantly lower than ΔPSE due to 20S). This means that the magnitude of virtual forearm *elongation* matters in determining the changes of TDP. On the other hand, the effect of *synchronicity* of the VHI on the modification of the TDP is absent. As expected, the significant difference in ΔPSE among the conditions of arm elongation means that the paradigm is able to modulate the tactile perception and, hence, the body schema. However, differently from studies performed in real environment (Van Der Hoort et al., 2011; Bruno and Bertamini, 2010), we found no correlation between RHI index and ΔPSE induced by VR.

The presence of the *elongation* effect and the absence of a significant *synchronicity* effect, combined also with the absence of correlation between RHI Index and ΔPSE suggest that in our study the modification of the TDP is not directly explained with the visuo-tactile integration elicited by the VHI paradigm. Although it is widely acknowledged that embodiment occurs through multisensory integration (Castro et al., 2023), the visual component (e.g., "visual capture phenomenon") (Pavani et al., 2000) alone may be sufficient to induce it; this effect in particular is reported in immersive 1PP virtual experience (Maselli and Slater, 2013; Tieri et al., 2015; Argelaguet et al., 2016; Bailey et al., 2016; Pavone et al., 2016; Spinelli et al., 2018; Bourdin et al., 2019; Matamala-Gomez et al., 2019). Here, the familiarization phase may have played a pivotal role, it is plausible that the sense of agency alone carried-over embodiment towards the virtual forearm (D'Angelo et al., 2018). In this perspective, given the strong impact of visual feedback on the subject's agency, visuo-tactile integration may not have played a primary role in incorporating the avatar's arm into the subjects' body representation yielding no effect of synchronicity. In line with this, a previous study found changes in the perceived size of objects after the exposure to VR environment: participants experienced, in 1PP VR, vision and agency over a virtual hand of modified dimensions but without any form of visuo-tactile stimulation (Linkenauger et al. 2013).

Although a significant effect of synchronicity was highlighted by the RHI index value, suggesting differences in embodiment levels between asynchronous and synchronous conditions, this does not necessarily indicate a disownership of the virtual hand in the asynchronous condition. In fact, many authors reported only reduced – but not null – illusion during the asynchronous condition (Ehrsson et al., 2004; Costantini & Haggard, 2007; Shimada et al.,2009). In addition, considering that the RHI questionnaire was an adapted version of the original one which was based on visuo-tactile integration (Botvinick and Cohen, 1998), the extracted index may not be capable to highlight the effect of visual capture on the avatar's embodiment.

As regards PD results, we found significant difference with respect to 0 of drift values for all conditions. Consistently with the aforementioned findings, the drift in perceived hand location may be mainly caused by the visual feedback from the elongated upper limb (Perez-Marcos et al., 2012). The absence of a virtual forearm elongation effect on the PD (no significant difference between 40S and 20S) is an intriguing finding. We could hypothesize that proprioception is malleable but bounded to some extent by our natural body schema (prior to its alteration) and insensitive to excessive distortions. Indeed, previous study found the PD to be significantly greater than 0 with a virtual forearm of two and three times the length the real limb length (equivalent to our 20cm and 40cm elongation respectively) but without any significant increase of the elicited PD for the times-three condition with respect to the times-two condition (Kilteni and al., 2012). They furthermore found the PD effect to be seemingly limited to a times-three forearm length by finding an absence of PD for a times-four forearm length.

Both TDP and PD seem to be affected by body schema modifications (De Vignemont, 2010; De Vignemont, 2011; Romano et al., 2015; Meraz et al., 2018). However, as shown by the significant effect of elongation, TDP seems to better highlight these changes. This may be due to the task resolution which makes TDP more suitable to measure perceptual changes, or, alternatively, it may disclose that exteroception and proprioception differently respond to our modulations.

In conclusion, in our everyday life, we interact with the environment using our body to sense. Like any sensor, our perception requires specific reference frames to measure parameters accurately. For instance, to perceive tactile distances we rescale our perception based on the currently perceived dimensions of the touched body part. It is believed that rescaling process involves the implicit model of the metric properties of our body constructed by the brain: the body schema. Several studies have proven the latter by modifying the body schema of participants through *own* body size modification illusions and observing a correlated modification of the TDP. However, to our knowledge, none had investigated it following a modification of the body schema through the *embodiment* of an artificial (i.e., virtual) body part. In this study, we investigated the effects on TDP of a virtual elongated forearm experienced in 1PP virtual environment. Even though we found an increase of the TDP positively associated with the virtual forearm *elongation* magnitude, no link has been found between the perceived embodiment over the virtually elongated arm resulting from the visuo-tactile synchronicity of the VHI and the perceived augmentation of the TDP, suggesting that the alteration of the body schema has taken place mainly through visual feedback over the elongated body part. We further hypothesized that a virtual embodiment due to the immersive 1PP experience of a virtual body might have taken place, dominating the VHI-induced embodiment.